\documentclass[aps,prl,reprint,groupedaddress,superscriptaddress]{revtex4-2}
\usepackage{graphicx,float}
\usepackage{amsmath}
\usepackage{amsfonts}
\usepackage{braket}
\usepackage{mathtools}
\usepackage{hyperref}
\usepackage{times,txfonts}
\usepackage[dvipsnames]{xcolor}
\DeclarePairedDelimiter\abs{\lvert}{\rvert}%

\begin{document}

\noindent\textbf{Comment on ``Quantum Limits to Incoherent Imaging are Achieved by Linear Interferometry''}

\bigskip

In a recent Letter \cite{lupo2020quantum}, Lupo {\it et al.}\ assert that linear interferometry followed by photon counting always saturates the quantum limit for imaging $N$ weak incoherent emitters. This conclusion rests entirely on the claim that there exists an optimal interferometer that saturates the Cauchy-Schwarz inequality $f_{r,r'}^c \geq f_{r,r'}$, where $f_{r,r'}^c$ and $f_{r,r'}$ are the classical and quantum fidelity, respectively. Here we show that the interferometer construction advertised in \cite{lupo2020quantum} (with full details reported in the Supplemental Material) does not, in general, saturate the Cauchy-Schwarz inequality. Consequently, the proof given in \cite{lupo2020quantum} does not establish the claimed saturation of the quantum limit for arbitrary source configurations. We identify where the argument breaks down and provide a corrected construction that exactly achieves the quantum limit, thereby restoring the main claim of \cite{lupo2020quantum} on rigorous foundations.

The error appears in the transition from Supplemental Eqs.~(31) to (32). Let $A' = RA$ and $B' = RB$ be purification matrices under the unitary transformation $R$. In Supplemental Eq.~(31), the authors correctly give the classical fidelity $f_{r,r'}^c$ as the sum of row norms:
\begin{equation}\label{eq1}
f_{r,r'}^c = \sum_{v} \|a'(v)\|\ \|b'(v)\|.
\end{equation}
They then use a QR decomposition to choose $R$ such that $A'$ is upper triangular and $B'$ is lower triangular. In Supplemental Eq.~(32), they claim (\ref{eq1}) simplifies to the sum of the products of the diagonal elements:
\begin{equation}\label{eq2}
f_{r,r'}^c \overset{?}{=} \sum_{v} \abs{a'(v,v)}\ \abs{b'(v,v)}.
\end{equation}

This simplification is generally invalid. For triangular matrices, $\|a'(v)\| \geq |a'(v,v)|$, with equality only if $A'$ and $B'$ are diagonal. Since a QR decomposition of an arbitrary matrix does not generally yield a diagonal form, the authors' construction can result in a classical fidelity strictly larger than the quantum fidelity ($f_{r,r'}^c > f_{r,r'}$).
This leads to a gap between the classical (CFI) and quantum Fisher information (QFI) for general source configurations, contradicting the main text \cite{lupo2020quantum}.

\begin{figure}[t!]
\includegraphics[width=0.87\columnwidth]{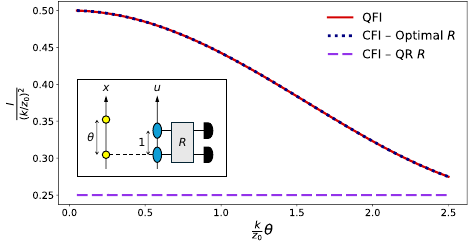}\vspace*{-0.2cm}
 \caption{Fisher information versus separation $\theta$ scaled by $k/z_0$, with sources at $x=0$ and $x=\theta$ along with collectors at $u=0, 1$ (inset), forming an asymmetric counterexample to the QR-based construction of \cite{lupo2020quantum}. Comparison between the quantum and classical curves shows suboptimality of the QR-based construction proposed in \cite{lupo2020quantum} and optimality of the interferometer $R$ presented here.}
    \label{fig1}
\end{figure}

Looking at the examples from \cite{lupo2020quantum}, both contain two point sources that are inversion-symmetric.
The symmetry ensures the singular vectors of $M=C(r)^\dagger C(r')$ diagonalise the Gram matrices $C(r)^\dagger C(r)$ and $C(r')^\dagger C(r')$, making $A$ and $B$ column-orthogonal and the QR decomposition diagonal, allowing (\ref{eq2}) to hold.
Breaking this symmetry immediately yields counterexamples; one such configuration is shown in Fig.~\ref{fig1}.

We now show how to correctly construct an optimal interferometer. Using $f_{r,r'} = \operatorname{Tr} (D) = \operatorname{Tr} (B^\dagger A) = \operatorname{Tr} (A B^\dagger)$, the quantum fidelity can be described as $f_{r,r'} = \sum_{v} \langle a'(v), b'(v) \rangle$, where $\langle a'(v), b'(v) \rangle = a'(v) b'(v)^\dagger$ are inner products on the $v^{\text{th}}$ rows of $A'$ and $B'$.
The saturation $f_{r,r'}^c = f_{r,r'}$ occurs when, for each pair of rows, either $a'(v)$ or $b'(v)$ vanishes or both are  proportional according to $b'(v) = \lambda_v a'(v), \,\lambda_v > 0$.

To enforce saturation $f_{r,r'}^c = f_{r,r'}$, decompose $A = [Q_S\,|\,Q_N] \left[ {{A_S \atop 0}}\right]$,
where $Q_S$ and $Q_N$ span the support and null space of the column space of $A$, respectively.
Let $P_S = B_S A_S^+$ be the unique solution to $B_S = P_S A_S$, where $B_S = Q_S^\dagger B$.
The diagonal matrix $A_S^\dagger B_S = D \geq 0$ shows that $P_S = (A_S^+)^\dagger D A_S^+$
is Hermitian and positive semi-definite. The diagonalisation $P_S = R_s^{\dagger}\Lambda R_S$ ensures row proportionality, $R_S B_S = \Lambda R_S A_S$, yielding the optimal interferometer $R = [R_S \oplus I] Q^\dagger$.

This construction attains the QFI. To show it, we expand $B_S = A_S + \delta \theta A_S^{(1)} + O(\delta \theta^2)$ giving
$P_S \approx I + \delta \theta X_S$, where $X_S = A_S^{(1)} A_S^+$ is the unique solution to $A_S^{(1)} = X_S A_S$. In the infinitesimal limit $\delta\theta \rightarrow 0$, $P_S \rightarrow I$ and  $R_S$  diagonalises $X_S$,
ensuring $R_S A_S^{(1)} = \Lambda_X R_S A_S$ with real $\Lambda_X = \text{diag}(\lambda_{X,v})$.
Using $b'(v) = a'(v) + \delta \theta a'^{(1)}(v) + \delta \theta^2 a'^{(2)}(v)$ and defining $S: =  \{v:a'(v) \neq 0\}$,
we find expressions
$\frac{\partial^2 f^c_{r,r'}}{\partial \delta \theta^2} = \sum_{v \in S} \left( \|a'^{(1)}(v)\|^2 + 2\text{Re}\langle a'(v), a'^{(2)}(v) \rangle - \frac{(\text{Re}\langle a'(v), a'^{(1)}(v) \rangle)^2}{\|a'(v)\|^2} \right)$ and $\frac{\partial^2 f_{r,r'}}{\partial \delta \theta^2} = 2 \sum_{v\in S} \operatorname{Re} \langle a'(v), a'^{(2)}(v) \rangle$ for the second derivatives of classical and quantum fidelity.
Taking their difference gives  $\frac{\partial f^c_{r,r'}}{\partial \delta\theta^2} - \frac{\partial f_{r,r'}}{\partial \delta\theta^2} =   \sum_{v \in S} \left( \|a'^{(1)}(v)\|^2  - \frac{(\text{Re}\langle a'(v), a'^{(1)}(v) \rangle)^2}{\|a'(v)\|^2} \right)$.
Since $R_S A_S^{(1)} = \Lambda_X R_S A_S$ implies $a'^{(1)}(v) = \lambda_{X,v} a'(v)$, this difference vanishes, proving that our construction  saturates the QFI.

With this construction of the optimal linear interferometer $R$ --- and not, in general, with the published QR-based configuration --- the main claim of the Letter~\cite{lupo2020quantum} can be restored.

\smallskip

\medskip

\noindent George Brumpton,$^{1}$ Aiman Khan,$^1$ Helia Hooshmand,$^1$ Samanta Piano,$^1$ and Gerardo Adesso$^{2}$

\hangindent=1.5em\hangafter=1
{\small{$^1$Manufacturing Metrology Team, Faculty of Engineering, University of Nottingham, Nottingham NG8 1BB, UK}}

\hangindent=1.5em\hangafter=1
{\small{$^2$School of Mathematical Sciences, University of Nottingham, University Park, Nottingham NG7 2RD, UK}}


%

\end{document}